# Nuclear Targets for a Precision Measurement of the Neutral Pion Radiative Width

P. Martel[j], E. Clinton[j], R. McWilliams[j], D. Lawrence[j], R. Miskimen[j], A. Ahmidouch[m], P. Ambrozewicz[m], A. Asratyan[a], K. Baker[e], L. Benton[m], A. Bernstein[k], P. Cole[f], P. Collins[b], D. Dale[f], S. Danagoulian[m], G. Davidenko[a], R. Demirchyan[m], A. Deur[g], A. Dolgolenko[a], G. Dzyubenko[a], A. Evdokimov[a], J. Feng[n], M. Gabrielyan[h], L. Gan[n], A. Gasparian[m], O. Glamazdin[i], V. Goryachev[a], V. Gyurjyan[g], K. Hardy[m], M. Ito[g], M. Khandaker[l], P. Kingsberry[l], A. Kolarkar[h], M. Konchatnyi[i], O. Korchin[i], W. Korsch[h], S. Kowalski[k], M. Kubantsev[o], V. Kubarovsky[g], I. Larin[a], V. Matveev[a], D. McNulty[k], B. Milbrath[p], R. Minehart[r], V. Mochalov[q], S. Mtingwa[m], I. Nakagawa[h], S. Overby[m], E. Pasyuk[b], M. Payen[m], R. Pedroni[m], Y. Prok[k], B. Ritchie[b], C. Salgado[l], A. Sitnikov[a], D. Sober[c], W. Stephens[r], A. Teymurazyan[h], J. Underwood[m], A. Vasiliev[q], V. Verebryusov[a], V. Vishnyakov[a], M. Wood[j]

*(a) Alikhanov Institute for Theoretical and Experimental Physics, (b) Arizona State University, (c) Catholic University, (d) Joint Institute for Nuclear Research, Dubna, (e) Hampton University, (f) Idaho State University, (g) Jefferson Lab, (h) University of Kentucky, (i) Kharkov Institute of Physics and Technology, (j) University of Massachusetts, Amherst, (k) Massachusetts Institute of Technology, (l) Norfolk State University, (m) North Carolina A&T State University, (n) University of North Carolina, Wilmington, (o) Northwestern University, (p) Pacific Northwest National Laboratory (q) Institute for High Energy Physics, Protvino, (r) University of Virginia*

Corresponding author: Rory Miskimen, miskimen@physics.umass.edu

**Abstract**

A technique is presented for precision measurements of the area densities, $\rho T$, of approximately 5% radiation length carbon and $^{208}$Pb targets used in an experiment at Jefferson Laboratory to measure the neutral pion radiative width. The precision obtained in the area density for the carbon target is ± 0.050%, and that obtained for the lead target through an x-ray attenuation technique is ± 0.43%.

## 1. Introduction

Photo- and electro-nuclear experiments performed at the high duty factor, high intensity electron accelerators such as Jefferson Laboratory and the MAMI Laboratory are continually pushing towards higher levels of precision. With the dramatic increase in beam intensity and beam quality available at these labs, and the availability of fully developed state-of-the-art detection systems, the relative importance in minimizing the uncertainty in the target area density, $\rho T$, has increased. Examples of experiments that require precise target information include measurements of elastic electron scattering form factors, and the quasi-free electron scattering response functions on nuclear targets.

Recently at Jefferson Lab, the PRIMEX collaboration [1] made a precision measurement of the neutral pion radiative width through a measurement of Primakoff cross sections on carbon and lead. A precision measurement of the neutral pion lifetime is regarded as one of the definitive tests of quantum chromodynamics at low energy, and the experimental value for the radiative width 7.84 ± 0.56 eV provides textbook confirmation for $N_c = 3$. Next-to-leading-order calculations [2], which include quark mass effects and $\pi^0$-$\eta$ mixing, are expected to be accurate at the percent level, while the Particle Data Group average for the $\pi^0$ radiative width has an error of 7%. The goal of the PRIMEX experiment is to reduce the error on the $\pi^0$ radiative width to the level where theories can be tested.

For the PRIMEX experiment to succeed it was necessary to measure absolute differential cross sections for coherent $\gamma A \rightarrow \pi^0 A$ with unparalleled statistical and systematic accuracy. Initial plans for the experiment had allowed for errors as large as ±0.7% in the target area densities. This article presents the techniques that allowed us to go under the budgeted 0.7% error by a factor of 14 for the carbon target, and a factor of 1.6 for the lead target.

2. **Carbon Target**

The carbon target was machined from a block of Highly Ordered/Oriented Pyrolitic Graphite (HOPG) obtained from SLAC. HOPG is produced using high temperature (3273 $^O$K) Chemical Vapor Deposition furnace technology, which creates atomic layers of carbon in an ordered pattern. An advantage of HOPG compared to normal graphite is the very low porosity of the HOPG, 1% as compared to graphite's 10% porosity.

The machined HOPG target block had dimensions of $.94 \times .94 \times .38$ in$^3$. A micrometer with precision of ±0.05 mils was used to map the thickness of the HOPG target in the 0.38 inch direction over the surface area of the target. The thickness was found to vary by approximately ±0.04% over the central 0.4 inch region of the target.

The mass density of the HOPG material was measured using the water immersion technique. HPLC grade $H_2O$ was used, which is submicron filtered, packed under inert gas, and has a maximum limit of impurities at 1 ppm. Corrections were made for the temperature dependence of the water density. A microgram scale was used to weight the target block in air and in the water. The mass densities of two identical HOPG blocks were measured three consecutive times, and the results are shown in Fig. 1. The mass density used for calculating the areal density of the target was the average of the first five measurements; trial #6 was excluded from the average. The error in mass density is taken from trial #3.

Two corrections are applied to $\rho T$ to obtain the effective area density for the target. The first correction accounts for impurities in the target, which can produce neutral pions through the Primakoff process. This effect was estimated using the known functional form of the Primakoff cross section on nuclear targets [3], and the known elastic electron scattering form factors [4]. Impurities in the HOPG material were determined using (i) Optimum Combustion Methodology, which detects C, H, N, and O, and (ii) PIXE analysis, which detects 72 heavier elements. The correction factor to the density was approximately +0.1%. Calculations [5] indicate that magnetic Primakoff production from $^{13}C$ is reduced by a factor of approximately $10^6$ compared to Coulomb Primakoff production. Therefore, the Primakoff cross sections on $^{13}C$ and $^{12}C$ can be treated as equal, and no correction is needed to account for the $^{13}C$ content in the natural isotopic HOPG target.

The second and larger correction accounts for the attenuation of the incident photon beam in the target. The NIST XCOM data base[6] was used to calculate the effect of incident beam absorption. The error in the attenuation coefficient is estimated [7] at 1.5%. The final result for the effective number of $^{12}C$ atoms/cm$^2$ at target center is $N_{eff}(^{12}C) = 1.0461 \times 10^{23}$ atoms/cm$^2$ ± 0.050%.

3. **$^{208}Pb$ Target**

The $^{208}Pb$ target, manufactured by Oak Ridge National Laboratory [8], is a one inch diameter circular foil of 99.09% enriched $^{208}Pb$, with an approximate thickness of 12 mil (1mil = .001 inch). Because the lead target is a thin, soft foil that can be easily bent and dented, it was decided to minimize the number of micrometer measurements over the surface of the target. The technique used is based on x-ray attenuation. X-rays from a 10 µCi activity $^{241}$Am source are collimated to a spot size of approximately 2 mm on the lead target, and the transmitted x-rays are detected in a 1" diameter NaI crystal coupled to a photomultiplier. Fig. 2 shows pulse height spectra taken with several different lead absorbers, ranging from no absorber up to 16.96 mil of Pb. Also shown is the background distribution with no x-ray source present. The prominent peak at the center of the spectrum is the 60 keV x-ray line from $^{241}$Am. Fig. 2 demonstrates the strong correlation between the strength of the transmitted 60 keV line and the lead absorber thickness, which is utilized to deduce the thickness of the target.

The transmitted 60 keV x-ray intensity can be represented by

$$I(T) = I_0 B(T,\lambda) e^{-T/\lambda} \tag{1}$$

where $I$ is the transmitted intensity, $I_0$ is the unattenuated intensity, $T$ is the foil thickness in cm, and $\lambda$ is the 60 keV x-ray attenuation length for lead in units of cm. $B(T,\lambda)$ is the buildup factor [9] which accounts for x-rays Compton scattering into the acceptance of the NaI detector, which depends critically on the specific experimental setup for the measurement. Measurements over a wide range of lead foil thicknesses showed that the build-up factor can be represented by

$$B(T,\lambda) = 1 + \alpha \frac{T}{\lambda} \tag{2}$$

where $\alpha$ is a dimensionless constant. Terms of order $O(T/\lambda)^2$ and higher were not significant.

To determine the value for $\alpha$ in Eqn. 2, micrometer calibration and x-ray absorption measurements were taken at four off-center points on the lead target. The calibration points were at $(x,y)$ coordinates (220, 220), (220,-220), (-220,220), and (-220,-220) relative to target center, in units of mil. The data show that $\alpha T/\lambda \ll 1$, with $\alpha T/\lambda = 0.0672$. Therefore, the T dependence in Eqn. (1) can be closely approximated by a pure exponential with an effective absorption length $\lambda'$ that includes the effects of x-ray multiple scattering,

$$I(T) = I_0 e^{-T/\lambda'} \tag{3}$$

Solving for the target thickness $T$ gives,

$$T = \lambda' \ln \frac{I_0}{I} \tag{4}$$

The technique used in this analysis was to fix $\lambda'$ from the micrometer and x-ray calibration points, then use Eqn. 4 and x-ray absorption data to map thickness over the surface of the target.

A *Labview* [10] data acquisition system was used to acquire ADC data, and to control *X-Y* stepper motors that scanned the lead target over the collimated $^{241}$Am source. The duration of one measurement at one spot on the target was typically two hours. A CAMAC charge integrating ADC was used to obtain pulse height spectra. To obtain x-ray yields, the distributions were integrated from the point where the no-absorber pulse-height distribution dropped to a minimum between the iodine escape peak and the 60 keV line (see Fig. 2), out to an energy where the 60 keV strength was negligible compared to background. The absolute x-y coordinates of the scanning system were fixed by scanning a 2 mm diameter iron rod over the $^{241}$Am source. When analyzed, the wire-scan data exhibit a deep trough in the transmitted x-ray intensity corresponding to the iron wire passing directly over the source. The fitted position of the trough gives the position of the wire in *(x,y)* stepper coordinates.

Fig. 2 shows that room backgrounds in the NaI detector were significant, and it was necessary to subtract these backgrounds from the transmitted and the unattenuated intensities. To reduce room backgrounds to an acceptable level, the NaI and attached PMT were placed at the center of an iron pipe with an inside diameter of 2.75 inches, an outside diameter of 6 inches, and a length of 12 inches. Without this shield, the room backgrounds in the NaI were approximately 20 times worse than the level shown in Fig. 2. A cosmic-ray veto scintillating paddle placed over the NaI detector had a negligible effect on the backgrounds.

Two *(x,y)* target scans were performed. The first scan was with a 200 mil step in *(x,y)*, and a second scan with a 100 mil step was run approximately one month later. Figure 3 shows the results of the study. For a given step size, the four measurements of the effective x-ray absorption length $\lambda'$ agree, although there is a shift in $\lambda'$ when comparing the results with the 200 mil step versus the 100 mil step size. To keep the analyses consistent, the 200 mil step effective absorption length was applied to the 200 mil step data, and the 100 mil step effective absorption length was applied to the 100 mil step data. The consistency of the analysis can be checked by verifying that the target thicknesses obtained in the two independent scans agree within errors.

Figs. 4 and 5 show thickness crosscuts through the lead target along the vertical (*Y*) and horizontal (*X*) axes, respectively. Results from the 100 mil and 200 mil *(x,y)* step scans are shown on the plots, and there is good agreement between the two data sets. The figures indicate a thickness plateau near the center of the target that extends out to a radius of approximately 200 mil. During data taking for PRIMEX, the target ladder was positioned so that the beam went through the center of the lead target to take advantage of the relatively uniform target thickness in this area of the target.

The mass density for natural isotopic lead used in this analysis, $11.336 \pm .007$ g/cm$^3$, is the average of the minimum and maximum density values listed in several standard references [11], with an uncertainty taken as one-half the difference between maximum and minimum densities. Isotopic corrections were applied to the natural isotopic lead mass density to obtain the mass density of the enriched target.

As in the case of the HOPG target, corrections were applied to account for the effects of impurities in the target, and for attenuation of the incident photon beam. The final result for the effective number of $^{208}$Pb atoms/cm$^2$ at target center is $N_{eff}(^{208}Pb) = 9.875 \times 10^{20} \pm 0.43\%$ atoms/cm$^2$.

4. **Summary and conclusions**

Photo- and electro-nuclear experiments at mature facilities such as Jefferson Laboratory and MAMI are trending towards higher statistical and systematic precision. A significant source of error in experiments that use solid nuclear targets is the target area density, $\rho T$. An example of this is the PRIMEX experiment at Jefferson Lab, which is a precision measurement of the neutral pion radiative width. The PRIMEX experiment uses carbon and $^{208}$Pb targets with an approximate thickness of 5% radiation length. We have presented a technique that allowed a measurement of $\rho T$ for a relatively thick carbon target to the ±0.050% level. For the $^{208}$Pb target, an x-ray attenuation method has been presented that minimized the amount of direct contact with the delicate lead foil. The uncertainty obtained for the lead target area density over the surface of the target is ±0.43%. The limiting factor in the lead measurement is the ±0.05 mil accuracy of the micrometer used for calibrating x-ray attenuation in the target. If higher precision can be obtained at a single calibration point on the target by some other measurement technique, then by obtaining larger statistics in the x-ray transmission spectra (i.e. longer data collection runs and/or by a more intense radioactive source) it will be possible to carry the higher precision from the single calibration point to all other points on the target.

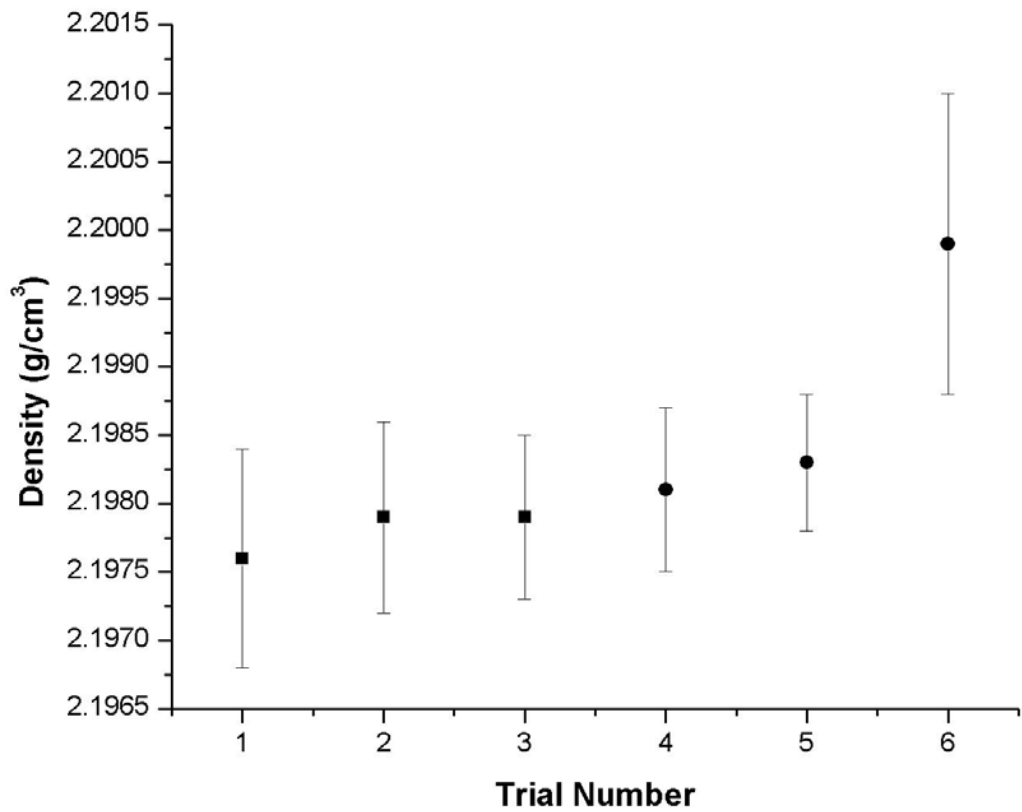

**Fig. 1.** Measurements of carbon target mass densities in units of g/cm$^3$. The square data points are from target block #1; the round data points are from target block #2.

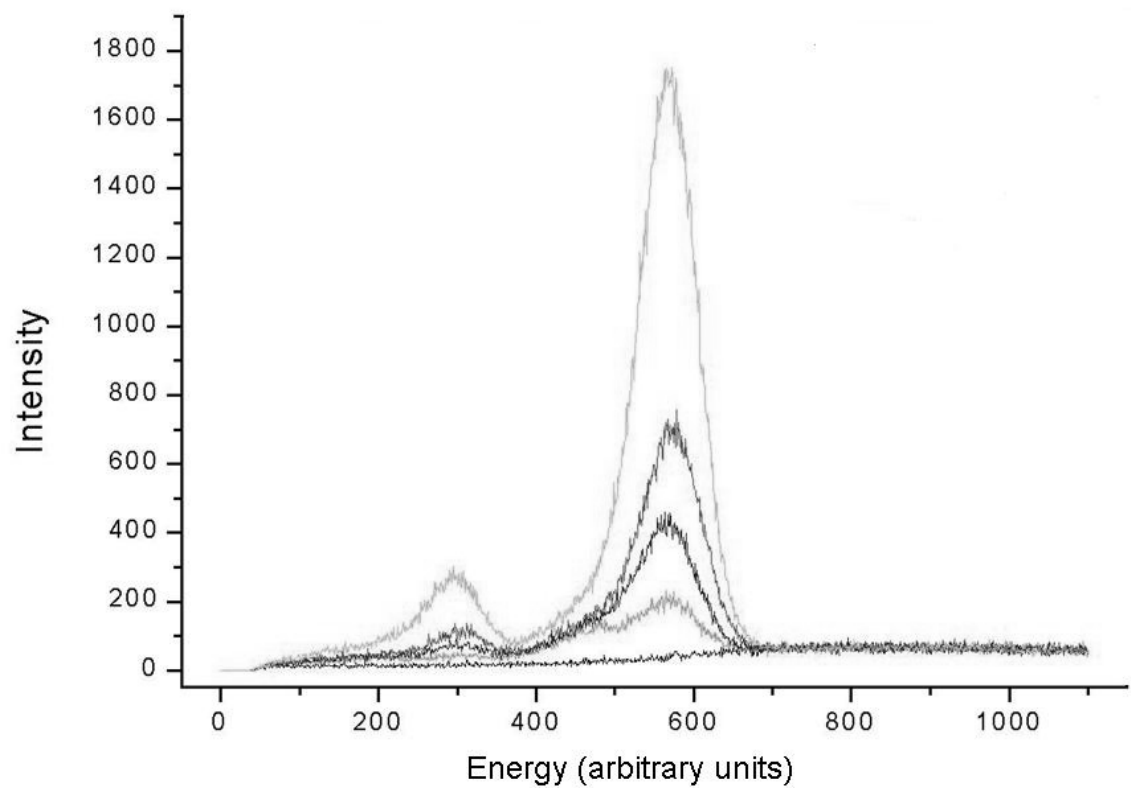

**Fig. 2.** Relative pulse height intensities for $^{241}$Am. In order from highest to lowest peak heights plotted in the graph, the spectra were taken with (a) no absorber, (b) a 6.95 mil lead absorber, (c) a 10.8 mil lead absorber, (d) a 16.95 mil lead absorber, and (e) with no source present (background spectrum).

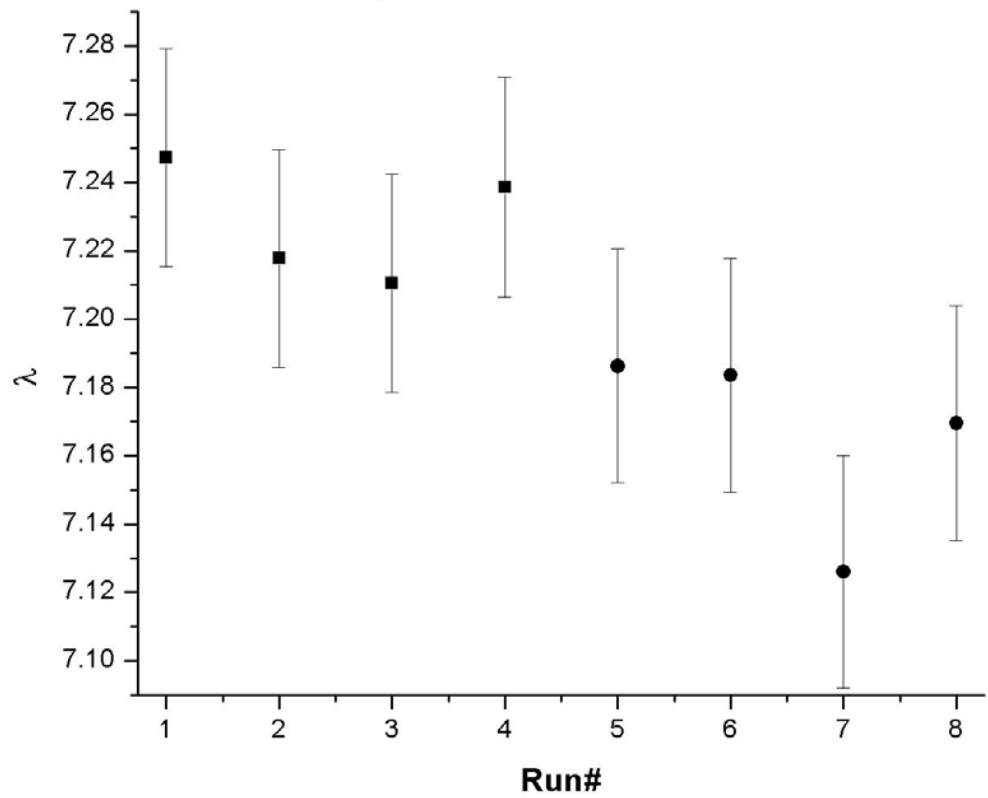

**Fig. 3.** Effective x-ray absorption lengths (λ) for the lead target in units of mil (.001 inch) for the 200 mil step size scan (square points), and the 100 mil step size scan (round points).

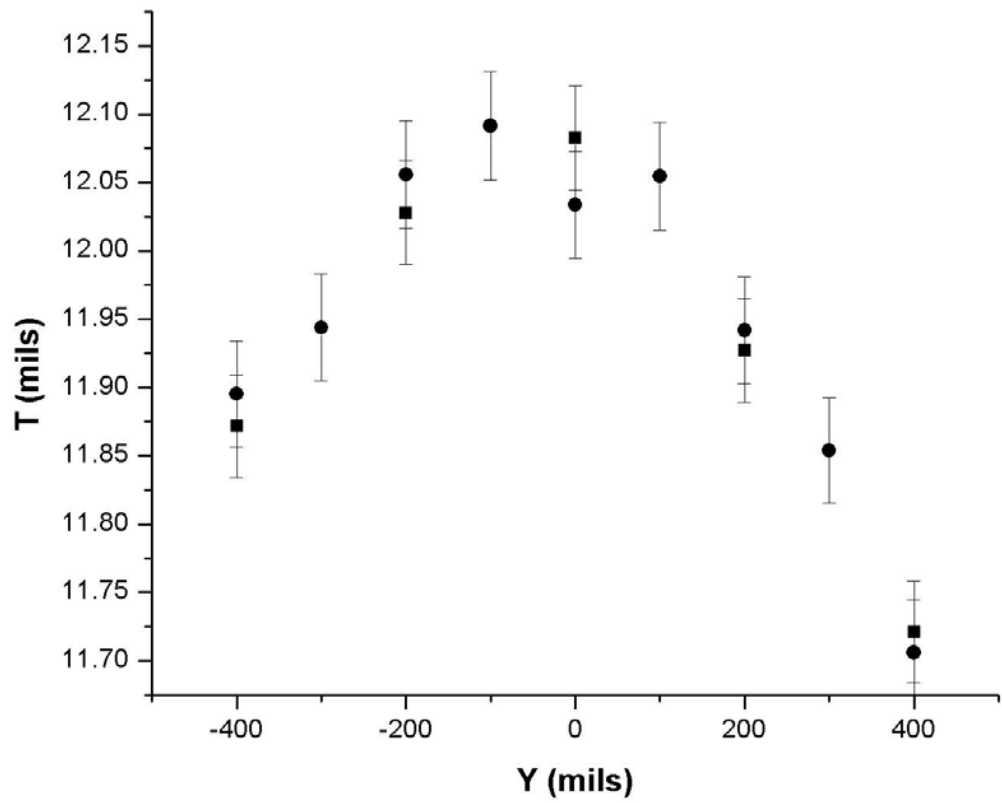

**Fig. 4.** Crosscut of the lead target thickness (T) along the vertical (Y) axis. The units of T and Y are mil. The square points were obtained from the 200 mil scan of the target; the round points were obtained from the 100 mil scan of the target.

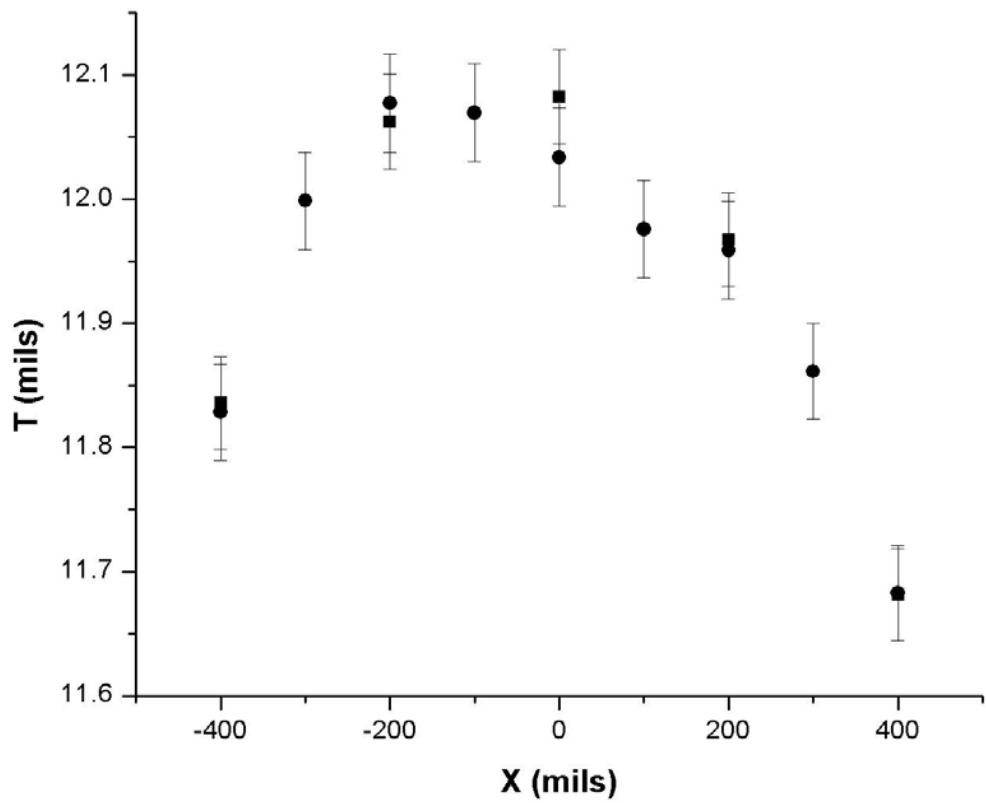

**Fig. 5.** Crosscut of the lead target thickness (T) along the horizontal (X) axis. The units of T and X are mil. The square points were obtained from the 200 mil scan of the target; the round points were obtained from the 100 mil scan of the target.